\documentclass[11pt,twoside]{article}

\usepackage{asp2006}
\usepackage{epsf}
\usepackage{psfig}
\usepackage{lscape}

\begin{document}
\title{Extended Commissioning and Calibration of the Dual-Beam Imaging Polarimeter} 
\author{Joseph Masiero, Klaus Hodapp, David Harrington, Haosheng Lin}
\affil{Institute for Astronomy, University of Hawaii, 2680 Woodlawn Dr, Honolulu, HI 96822, {\it masiero, hodapp, dmh, lin@ifa.hawaii.edu}}

\begin{abstract} 
In our previous paper \citep{JRM_dbip} we presented the design and initial
calibrations of the Dual-Beam Imaging Polarimeter (DBIP), a new optical
instrument for the University of Hawaii's $2.2~$m telescope on the summit of
Mauna Kea, Hawaii.  In this followup work we discuss our full-Stokes mode
commissioning including crosstalk determination and our typical observing
methodology.
\end{abstract}


\section{Introduction}
To study the linear polarization of asteroids and other point source objects,
the Dual-Beam Imaging Polarimeter (DBIP) was commissioned in March of 2007
\citep{JRM_dbip}.  In August of 2007 we expanded DBIP's capabilities to
include analysis of circular polarization with the addition of a quarterwave
plate.  Typically, the most important quantities for analysis are the
fractional polarizations $q=Q/I$, $u=U/I$, and $v=V/I$, expressed as
percentages, and in the following text we will deal with these quantities
when we refer to polarization measurements.  Here we present our subsequent
calibration and determination of systematic errors which were found to be
comparable to statistical errors for typical observing situations:
$\sim0.1\%$ polarization.

\section{Optical Setup}
\begin{figure}[h]
\plotfiddle{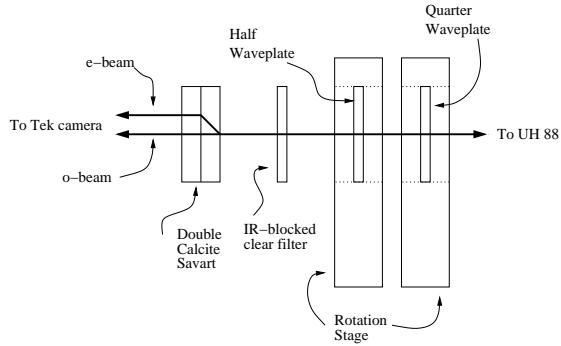}{1.75in}{0}{40}{40}{-100}{0}
\caption{Schematic of the optical components of DBIP.  Modified from \citet{JRM_dbip}.}
\label{JRM_fig.optics}
\end{figure}

The original setup of DBIP was a serial arrangement of a halfwave plate in an
encoded rotation stage, a filter and a double-calcite Savart plate placed
between the telescope and the $2k \times 2k$ Tektronix CCD camera.  To extend
DBIP to full-Stokes sensitivity, a quarterwave plate in a rotation stage was
placed ahead of the halfwave plate.  This setup allows for simultaneous
measurement of linear and circular polarization, though at the potential cost
of increased crosstalk between polarizations, which is discussed further in
\S\ref{JRM_crosstalk}  Figure~\ref{JRM_fig.optics}, modified from \citet{JRM_dbip},
shows a schematic representation of the new optical path with the
quarterwave plate added.

\section{Calibration}
As with any optical system, misalignments and imperfections in the components
will lead to errors in measurement.  In the case of DBIP, the waveplates are
the most sensitive components to these errors, as they are the only moving
parts and require precisely determined angular zero-points.  Errors in
angular alignment of the waveplate or tilt with respect to the optical axis
as well as chromatic retardance or fast-axis angle variations will show up in
our system as variations in measured position angle of polarization,
depolarization of the signal, or crosstalk between linear and circular
polarization.  To minimize and quantify these errors we performed an
extensive calibration campaign.

\subsection{Waveplate Alignment}
\label{JRM_align}
Our first step of calibration was to determine the alignment of the
waveplate zero-points using known standard stars.  Having already aligned
the halfwave plate against standards before the installation of the
quarterwave plate \citep{JRM_dbip}, we were able to re-observe one of the same
polarization standards (NGC 2024-1) in full-Stokes mode to align the
quarterwave zero-point while confirming that we could reproduce the linear
polarization results for this target.  The set of observations of NGC 2024-1,
both before and after the addition of the quarterwave plate, are listed in
Table~\ref{JRM_tab.stds}, where a circular polarization value of ``---''
indicates a measurement taken before the installation of the quarterwave
plate.

\begin{table}[h]
\caption{Polarized Standard Star Observations}
\smallskip
\begin{center}
{\scriptsize
\begin{tabular}{ccccccc}
\tableline
\noalign{\smallskip}
Name & Obs Date & $\%~$Lin Pol$_{lit}$ & $\theta_{lit}$ & $\%~$Lin Pol$_{obs}$ & $\theta_{obs}$ & $\%~$Circ Pol$_{obs}$ \\
\noalign{\smallskip}
\tableline
\noalign{\smallskip}
BD-12 5133 & 3/24/07 & $4.37 \pm 0.04$ & $146.84 \pm 0.25$ & $4.26 \pm 0.01$ & $146.20 \pm 0.10$ &  --- \\
\tableline
NGC 2024-1 & 3/24/07 & $9.65 \pm 0.06$ & $135.47 \pm 0.59$ & $9.70 \pm 0.02$ & $136.11 \pm 0.05$ &  --- \\
NGC 2024-1 & 1/17/08 & $9.65 \pm 0.06$ & $135.47 \pm 0.59$ & $9.68 \pm 0.04$ & $135.78 \pm 0.12$ & $-0.04 \pm 0.04$\\
NGC 2024-1 & 3/12/08 & $9.65 \pm 0.06$ & $135.47 \pm 0.59$ & $9.65 \pm 0.02$ & $135.75 \pm 0.05$ & $ 0.12 \pm 0.02$\\
\tableline
BD-13 5073 & 5/14/08 & $3.66 \pm 0.02$ & $152.55 \pm 0.11$ & $4.62 \pm 0.05$ & $149.28 \pm 0.31$ & $ 0.07 \pm 0.05$\\
\tableline
BD-12 5133 & 5/14/08 & $4.37 \pm 0.04$ & $146.84 \pm 0.25$ & $4.30 \pm 0.05$ & $146.26 \pm 0.31$ & $-0.04 \pm 0.05$\\
BD-12 5133 & 6/11/08 & $4.37 \pm 0.04$ & $146.84 \pm 0.25$ & $4.29 \pm 0.03$ & $145.17 \pm 0.21$ & $ 0.02 \pm 0.03$\\
\tableline
VI Cyg 12   & 6/11/08 & $8.95 \pm 0.09$ & $115.00 \pm 0.30$ & $8.69 \pm 0.04$ & $116.04 \pm 0.13$ & $-0.16 \pm 0.04$\\
\noalign{\smallskip}
\tableline
\end{tabular}
}
\label{JRM_tab.stds}
\end{center}
\end{table}

\begin{table}[h]
\caption{Unpolarized Standard Star Observations}
\smallskip
\begin{center}
{\scriptsize
\begin{tabular}{cccccc}
\tableline
\noalign{\smallskip}
Name & Obs Date & $\%~$Lin Pol$_{lit}$ & $\%~$Lin Pol$_{obs}$ & $\theta_{obs}$ & $\%~$Circ Pol$_{obs}$ \\
\noalign{\smallskip}
\tableline
\noalign{\smallskip}
HD 64299    & 03/23/07 & $0.15 \pm 0.03$ & $0.10 \pm 0.01$ & $83 \pm 4 $  & ---  \\
\tableline
WD 1615-154 & 03/24/07 & $0.05 \pm 0.03$ & $0.02 \pm 0.02$ & --- & ---  \\
WD 1615-154 & 03/12/08 & $0.05 \pm 0.03$ & $0.04 \pm 0.03$ & --- & $-0.01 \pm 0.03$ \\
WD 1615-154 & 05/14/08 & $0.05 \pm 0.03$ & $0.02 \pm 0.03$ & --- & $0.08 \pm 0.03$ \\
WD 1615-154 & 06/11/08 & $0.05 \pm 0.03$ & $0.19 \pm 0.06$ & $158 \pm 9 $ & $0.05 \pm 0.06$ \\
\tableline
BD+28d4211 & 08/29/07 & $0.05 \pm 0.03$ & $0.02 \pm 0.02$ & --- & $0.00 \pm 0.02$ \\
\tableline
WD 2149+021 & 08/30/07 & $0.04 \pm 0.01$ & $0.03 \pm 0.02$ & --- & $0.00 \pm 0.02$ \\
\tableline
G191B2B    & 01/17/08 & $0.06 \pm 0.04$ & $0.02 \pm 0.03$ & --- & $0.01 \pm 0.03$ \\
\noalign{\smallskip}
\tableline
\end{tabular}
}
\label{JRM_tab.unpol}
\end{center}
\end{table}

\subsection{Instrumental Polarization}
In order to test for instrumental polarization or depolarization, we have
observed polarized and unpolarized standard stars over a $15$ month baseline.
Tables~\ref{JRM_tab.stds} and \ref{JRM_tab.unpol} give our measured polarizations and
position angles for polarized and unpolarized standard stars, respectively,
as well as literature values for these objects from \citet{JRM_fossati07},
\citet{JRM_hubbleSTD2} and the Keck/LRISp
standards\footnote{http://www2.keck.hawaii.edu/inst/lris/polarimeter/polarimeter.html}.
Our measurements for both polarized and unpolarized standards agree within
$3~\sigma$ of the literature values, confirming that instrument systematics
are less than a $0.1\%$ effect.  The only exceptions to this are the
observations of BD-13 5073 and WD 1615-154.  BD-13 5073 clearly shows
evidence of variation in the amplitude and direction of polarization from the
literature values over only a few years, showing it cannot be depended upon
as a polarized standard.  Our observation of WD 1615-154 on 6/11/08 shows
anomalously high polarization compared to literature values and our previous
observations at the $\sim3~\sigma$ level.  With the current data it is
unclear if the polarization properties of the object have changed or if this
measurement is just an outlier.

\subsection{Crosstalk}
\label{JRM_crosstalk}
Instrumental crosstalk between Stokes vectors is one of the more subtle
errors that can affect polarization measurements and quantifying its
magnitude is a critical step toward obtaining high precision polarimetry.
Crosstalk between linear Stokes vectors ($Q$ to $U$ or $U$ to $Q$) happens
when the zero-point location of the halfwave retarder is offset from the
defined $Q$ direction, and is easily corrected by aligning the waveplate, as
discussed above in \S\ref{JRM_align} Crosstalk from circular to linear and back
($V$ to $Q\&U$ or $Q\&U$ to $V$) can be caused by waveplate offsets or by
variations in retardance of the wave-plates as a function of wavelength,
position on the plate, etc.

To fully determine the instrumental crosstalk, light with a known
polarization must be sent through the instrument and measured upon exiting.
It is possible to do this by installing polarizers just in front of the
instrument during observations \citep[recent examples include][etc.]{JRM_perrin08,JRM_snik06}, however
due to limited observing time and space in the optical path this was not a
feasible calibration method for DBIP.  Instead, we performed lab bench
calibrations to quantify instrumental crosstalk.

To measure $Q\&U$ to $V$ crosstalk we sent a collimated light source through a
linear polarizer into DBIP and measured one of the two exiting beams with a
precision photodiode.  By stepping the angular position of the input linear
polarizer, we were able to fully characterize this linear-to-circular
crosstalk.  To measure $V$ to $Q\&U$ crosstalk a fixed quarterwave plate was
installed after the linear polarizer.  By stepping the linear polarizer
again, we were able to vary the input polarization from circular to linear
and back and characterize this circular-to-linear crosstalk.

\begin{figure}[h]
\plotfiddle{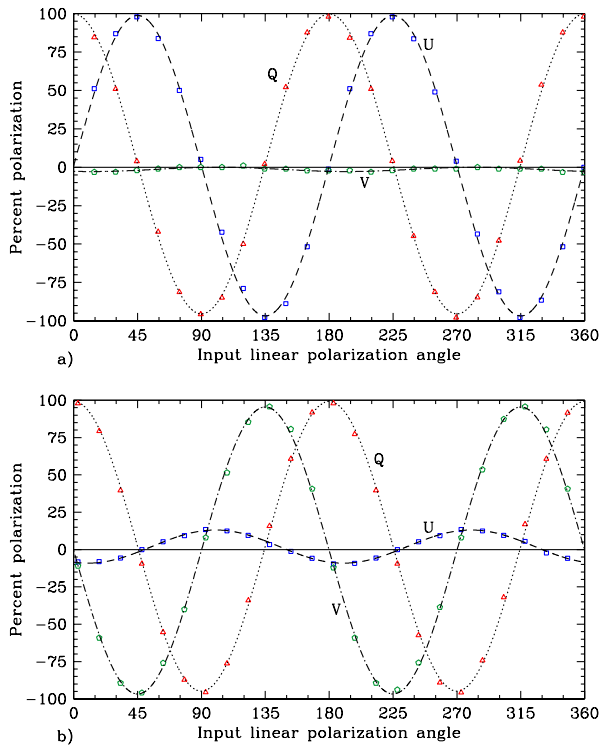}{4in}{0}{40}{40}{-110}{0}
\caption{Polarization crosstalk measurements for DBIP for two cases: (a) for
input of pure linear polarization rotated through 360 degrees, and (b) for
input of rotated linear polarization passed through a fixed quarterwave
plate.  Sinusoids for both cases were fitted using a chi-squared minimizer,
and are labeled with the Stokes vectors they represent. }
\label{JRM_fig.crosstalk}
\end{figure}

Figure~\ref{JRM_fig.crosstalk} shows our measured polarization state for both
setups as the input linear polarizer was stepped through $360~$degrees.
Sinusoids describe the behavior of the polarization vectors in both setups
and were fit with a chi-squared minimizer to the data to determine percentage
crosstalk as well as any depolarization from the optics.  We find that the
circular polarization measured when inputing pure linearly polarized light
had an amplitude of $1.4\%$ and an offset of $-1.4\%$, with crosstalk
preferentially occurring when both Q and U are positive (around $22^\circ$ on
our plot).  We measure a depolarization of both linear signals varying from
$<0.5\%$ near $22^\circ$ to $\sim4\%$ near $112^\circ$.  Since these
crosstalk errors are only a few percent of the input signal, for sources with
``typical'' polarizations (i.e. $5-10\%$) the crosstalk error is comparable
to the desired statistical errors of $\sim0.1\%$ polarization, as we measured
for our polarized and unpolarized standards.

For the measurement of circular-to-linear crosstalk, we find a small amount
of linear $u$ polarization generated from a purely circularly polarized input
beam, preferentially when $+v$ is input.  Note that the phased oscillation of
$q$ and $u$ indicate a small misalignment of the quarterwave plate axes with
the instrument-defined $Q$ orientation, while it is the slight phase offset
between $-u$ and $+q$ that is produced by the crosstalk.  We also see clear
evidence for a depolarization of the circularly polarized signal at the level
of $\sim4\%$ of the input polarization.  For targets with total circular
polarizations less than a few percent these errors should be comparable the
$0.1\%$ noise error typically obtained in our observations.

\section{Conclusion}
We have presented the results of our extended commissioning of DBIP into
full-Stokes mode.  Crosstalk values are shown to be small and can be ignored
for objects with ``typical'' (i.e. $<5\%$) polarizations.  With errors well
constrained DBIP is now available for scientific studies of linear and
circular polarization of point sources in the optical.  DBIP has already been
successfully used to characterize the polarization-phase curves of asteroids
with some of the most pristine compositions in the inner solar system
\citep{JRM_aquitania} and a number of other programs to study small solar
system bodies are currently under way.

\acknowledgements 
JM would like to thank Robert Jedicke and Colin Aspin for providing funding
to attend the Astronomical Polarimetry 2008 conference.  JM was partially
funded under NASA PAST grant NNG06GI46G.  The authors wish to recognize and
acknowledge the very significant cultural role and reverence that the summit
on Mauna Kea has always had within the indigenous Hawaiian community.  We are
most fortunate to have the opportunity to work with telescopes located on
this sacred mountain.


\begin{thebibliography}{}

\bibitem[Fossati et al.(2007)]{JRM_fossati07}
Fossati, L., Bagnulo, S., Mason, E., Landi Del'Innocenti, E., 2007, ASP Conf., 364, 503

\bibitem[Masiero \& Cellino(2008)]{JRM_aquitania}
Masiero, J. \& Cellino, A., 2008, Submitted to Icarus

\bibitem[Masiero et al.(2007)]{JRM_dbip}
Masiero, J., Hodapp, K.-W., Harrington, D. \& Lin, H., 2007, PASP, 119, 1126

\bibitem[Perrin, Graham,\& Lloyd(2008)]{JRM_perrin08}
Perrin, M.D., Graham, J.R., \& Lloyd, J.P., 2008, PASP, 120, 555

\bibitem[Schmidt, Elston, \& Lupie(1992)]{JRM_hubbleSTD2}
Schmidt, G.D., Elston, R., \&  Lupie, O.L., 1992, AJ, 104, 1563

\bibitem[Snik et al.(2006)]{JRM_snik06}
Snik, F., Bettonvil, F.C.M., Jagers, A.P.L., Hammerschlag, R.H., Rutten, R.J. \& Keller, C.U., 2006, ASP Conf., 358, 205

\end{thebibliography}
\end{document}